\documentclass[twocolumn,tighten]{aastex631}

\newcommand{\sm}{\textsubscript{\(\odot\)}}

\newcommand{\bvf}{Brunt--V\"ais\"al\"a}

\newcommand{\msol}{\ensuremath{\mathrm{M}_\odot} }

\usepackage{amsmath} 
\shortauthors{Ratnasingam et al.}
\graphicspath{{./}{figures/}}

\begin{document}

\title{On the Geometry of the Near-Core Magnetic Field in Massive Stars}

\author[0000-0002-7250-6524]{R.~P. Ratnasingam}
\affiliation{School of Mathematics, Statistics and Physics, Newcastle University, Newcastle upon Tyne, NE1 7RU, UK}

\author[0000-0001-7019-9578]{P.~V.~F. Edelmann}
\affiliation{Computer, Computational and Statistical Sciences (CCS) Division and Center for Theoretical Astrophysics (CTA), Los Alamos National Laboratory, Los Alamos, NM 87545, USA}

\author[0000-0001-7402-3852]{D.~M. Bowman}
\affiliation{School of Mathematics, Statistics and Physics, Newcastle University, Newcastle upon Tyne, NE1 7RU, UK}
\affiliation{Institute of Astronomy, KU Leuven, Celestijnenlaan 200D, 3001 Leuven, Belgium}

\author[0000-0002-2306-1362]{T.~M. Rogers}
\affiliation{School of Mathematics, Statistics and Physics, Newcastle University, Newcastle upon Tyne, NE1 7RU, UK}
\affiliation{Planetary Science Institute, Tucson, AZ 85721, USA}




\begin{abstract}
It is well-known that the cores of massive stars sustain a stellar dynamo with a complex magnetic field configuration. However, the same cannot be said for the field's strength and geometry at the convective-radiative boundary, which are crucial when performing asteroseismic inference. In this Letter, we present three-dimensional (3D) magnetohydrodynamic (MHD) simulations of a 7~\msol mid-main sequence star, with particular attention given to the convective-radiative boundary in the near-core region. Our  simulations reveal that the toroidal magnetic field is significantly stronger than the poloidal field in this region, contrary to recent assumptions. Moreover, the rotational shear layer, also important for asteroseismic inference, is specifically confined within the extent of the \bvf{} frequency peak. These results, which are based on the inferred properties of HD~43317, have widespread implications for asteroseismic studies of rotation, mixing and magnetism in stars. While we expect our results to be broadly applicable  across stars with similar \bvf{} frequency profiles and stellar masses, we also expect the MHD parameters (e.g. Re$_{\mathrm{m}}$) and the initial stellar rotation rate to impact the geometry of the field and differential rotation at the convective-radiative interface.

\end{abstract}



\section{Introduction} \label{sec:intro}

Stars born with masses above about 1.2 M\sm{} have a convective core and radiative envelope during the main sequence, with the fluid motions in the core exhibiting turbulent convection. This process sets up a magnetic dynamo in the core, the strength of which is influenced mainly by rotation rate. The study of the stellar dynamos has been extensive, both analytically \citep{PittsTayler1985,Spruit1999,Zahn2007,Auguston2019,Charbonneau2023} and numerically \citep{Brun2005,Zahn2007,Browning2008,Featherstone2009,Featherstone2011,Auguston2016}.   

Currently, there are a number of uncertainties in stellar structure theory to be resolved. Among these are the mechanisms responsible for setting stars' rotation and chemical mixing profiles, and interior magnetic fields. One particularly important uncertainty is the amount of convective-boundary mixing (CBM) at the interface between convective cores and radiative envelopes in main-sequence massive stars, which can provide fresh hydrogen to the nuclear-burning core and extend the main-sequence lifetime by upwards of 25\% \citep{Bowman2020c}. CBM and more generally interior rotation and mixing processes are controlled by largely uncalibrated prescriptions in one-dimensional (1D) stellar evolution models. Whereas, magnetic fields are often neglected entirely owing to their complexity, and yet have been inferred indirectly to impact the amount of CBM \citep{Briquet2012}. Only recently have we been able to diagnose interior magnetic fields thanks to the study of stellar pulsations --- asteroseismology \citep{aerts2010asteroseismology}.

There has been a great deal of advancement in understanding the rotation and mixing profiles of stars thanks to asteroseismology \citep{Aerts2021}. The high-radial order gravity modes in slowly pulsating B-type (SPB; $3 \lesssim M \lesssim 8$~M$_{\odot}$) stars are particularly sensitive to the \bvf{} peak just outside the convective core, and reveal precise core masses and near-core rotation rates \citep{Moravveji2015b, Moravveji2016b, Papics2017a, Szewczuk2021a, Michielsen2021a, Pedersen2021a}. However, asteroseismic studies of SPB stars have generally assumed rigid interior rotation profiles, and have also not considered the impact of an interior magnetic field. The exception to this is HD~43317 which is the only confirmed magnetic SPB star to have undergone forward asteroseismic modelling \citep{Papics2012a, Briquet2013, Buysschaert2017b, Buysschaert2018c}. Recently, \citet{Lecoanet2022a} used the best-fitting asteroseismic model of \citet{Buysschaert2018c} and {\sc dedalus} \citep{Burns2020} to solve the linear eigenvalue problem assuming a purely dipolar interior magnetic field geometry to constrain an upper limit of the near-core field strength of HD~43317 as approximately $5 \times 10^5$~G. This result, whilst being an important step forward in the novel field of magneto-asteroseismology \citep{Bowman2023b}, assumed the magnetic field geometry and a rigid radial rotation profile, which are simplifications if compared to previous numerical simulations. In this Letter, we perform 3D MHD simulations of a 7-M$_{\odot}$ mid-main sequence star and reveal the magnetic field geometry and rotational profile at the core-envelope boundary in such massive stars.

\section{Numerical setup} \label{sec:Numerical_setup}
Our 3D spherical simulations are done using {\sc RAYLEIGH} \citep{featherstone2022paper,featherstone_et_al_2022,Matsui_etal_2016}, an open-source pseudospectral code which solves the MHD equations in the anelastic approximation. The discretisation is done with a finite-difference scheme in the radial direction and spherical harmonics in the horizontal directions. The whole simulation domain is resolved with 2400 radial ($r$) grid points, 256 grid points in the polar ($\theta$) direction and 512 grids in the azimuthal ($\phi$) direction, equivalent to a spherical harmonic $\ell_{\rm max}=170$. The code implements parallel frameworks using Message Passing Interface (MPI) and Open Multi-Processing (OpenMP) to achieve efficient scaling to more than 10000 cores.  The simulation solves the MHD equations:

\begin{figure}[ht!]
        \centering
        \includegraphics[trim={0.0cm 0.0cm 0 0.0cm},clip,width=\columnwidth]{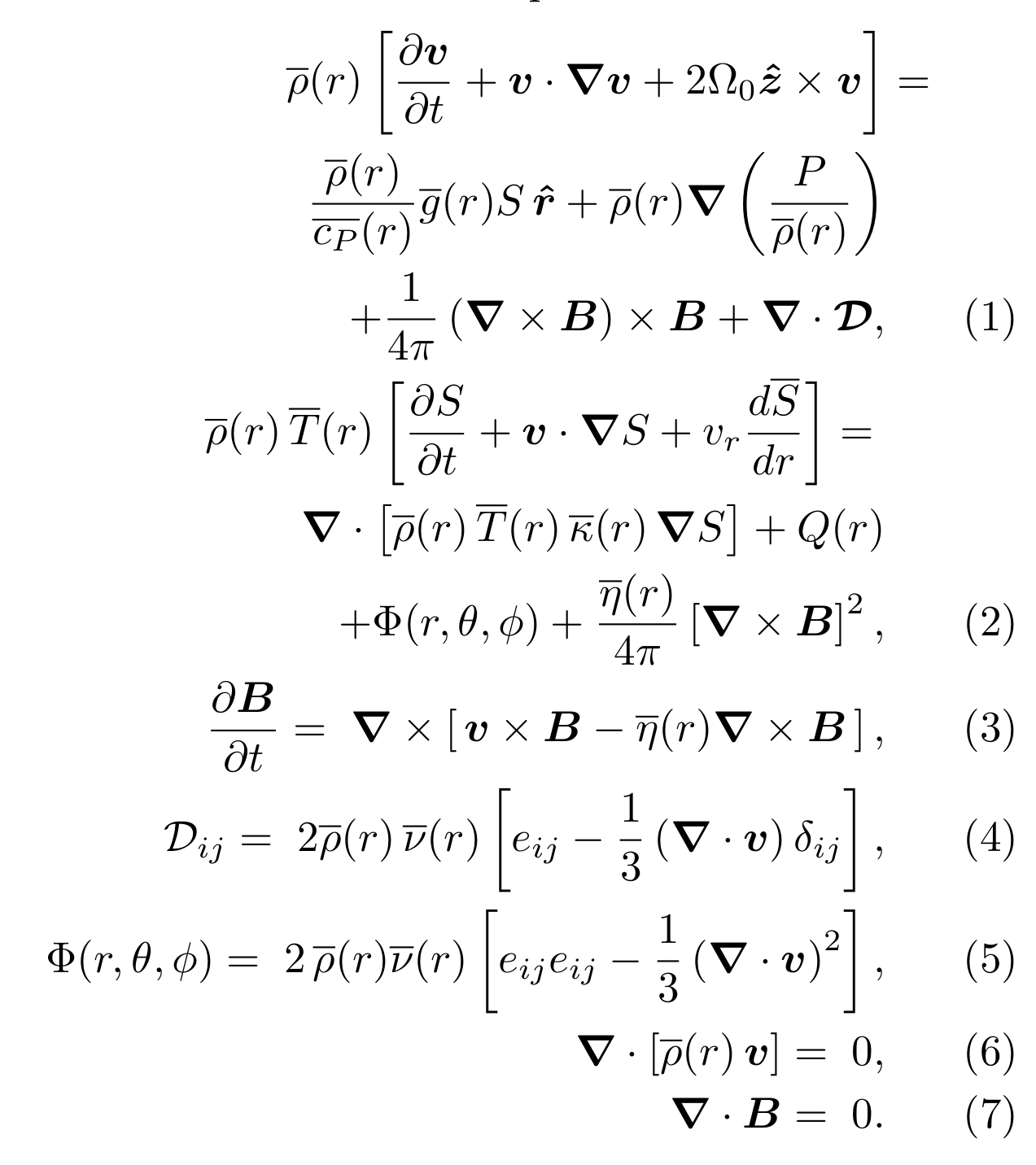}
        \centering
\end{figure}

The radial velocity, $v_r$, together with the latitudinal velocity, $v_{\theta}$, and the azimuthal velocity, $v_{\phi}$, form the velocity vector, $\boldsymbol{v}$. The radial, polar and azimuthal magnetic fields, represented by $B_r$, $B_{\theta}$ and $B_{\phi}$ respectively, form the magnetic field vector, $\boldsymbol{B}$. $S$ and $P$ are the perturbation entropy and pressure, respectively. The quantities with an overline are the radially dependent reference state values density, $\overline{\rho}$, specific pressure heat capacity, $\overline{c_P}$, gravity, $\overline{g}$, temperature, $\overline{T}$, entropy, $\overline{S}$, thermal diffusivity, $\overline{\kappa}$, magnetic diffusivity, $\overline{\eta}$, and viscosity, $\overline{\nu}$. 

The reference input state data are taken from the 1D stellar evolution code Modules for Stellar Astrophysics (MESA; \citealt{mesa_1,mesa_2,mesa_3,mesa_4,mesa_5}; r23.05.01) for an input model of a 7~M\sm{} star in the middle of the main sequence. The middle of the main-sequence here is defined to be when the central hydrogen mass fraction of the evolving star is $X_{\rm c} = 0.35$. For the MESA simulations, we set the stellar metallicity, $Z$, to be equal to the solar value of $Z$ = 0.02, the mixing-length parameter to be 1.8 and the convective overshoot profile is set to exponential (cf. \citealt{Buysschaert2018c}). Lastly, the stellar rotation rate, $\Omega_0$, was initially set to a solid body rotation rate of $1.8 \times 10^{-5}$ rad s$^{-1}$ (i.e. 4.04~d/11.5\% critical breakup velocity), which is slower than that of HD~43317. 

\begin{figure}[ht!]
        \centering
        \includegraphics[trim={0.0cm 0.0cm 0 0.0cm},clip,width=\columnwidth]{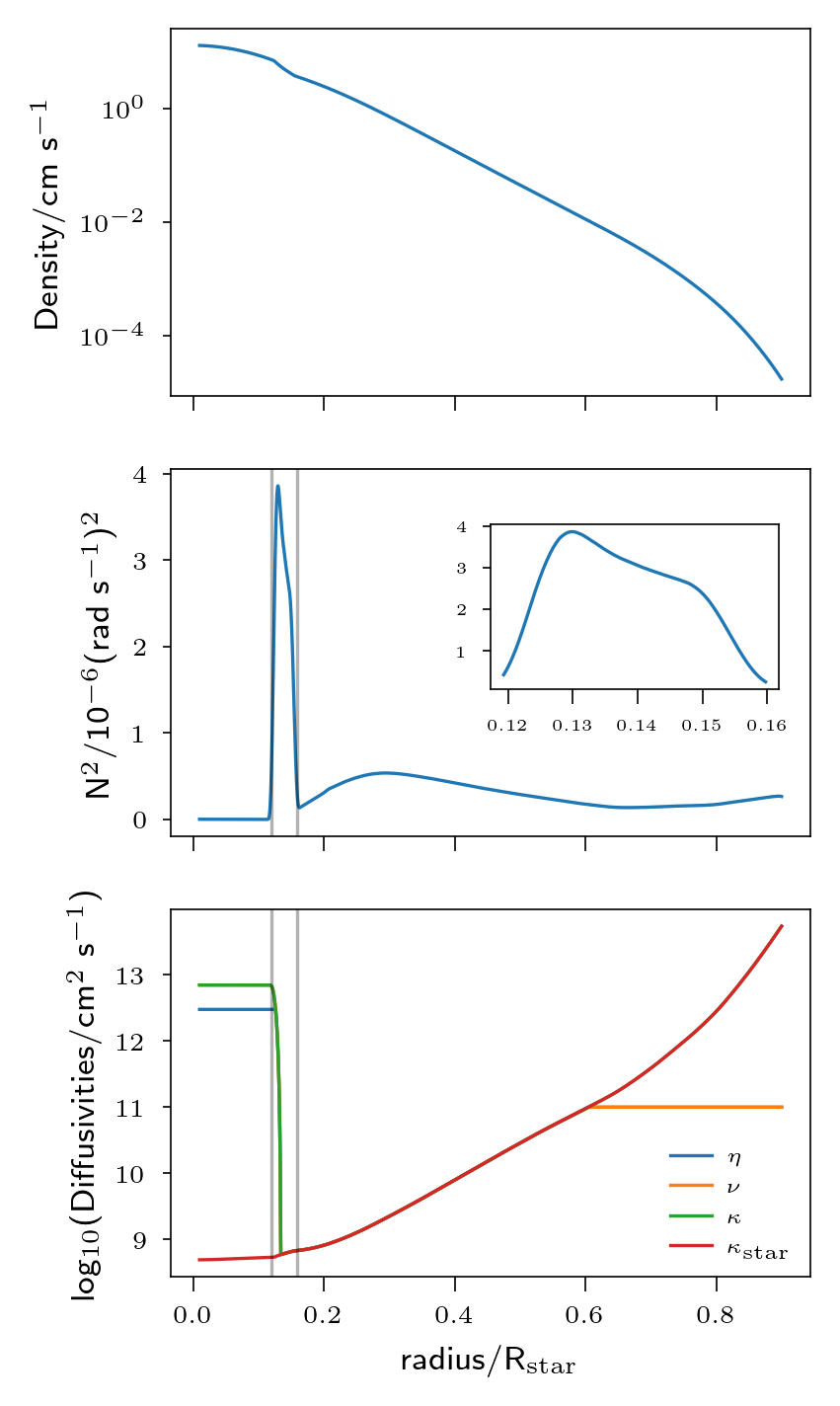}
        \caption{Reference state density, $\overline{\rho}$ (top panel), \bvf{} frequency squared (middle panel) and various diffusivity profiles (bottom panel) as functions of stellar radius in units of total stellar radius, R$_{\mathrm{star}}$. The solid, grey vertical lines show the extent of the \bvf{} frequency peak and the inset shows a zoom-in of this peak. The viscosity profile coincides exactly with the thermal diffusivity profile up to 0.6 R$_{\mathrm{star}}$. \label{fig:stellar_param}}
        \centering
\end{figure}

Figure~\ref{fig:stellar_param} shows the smoothed \bvf{} frequency, thermal diffusivity profiles and density of the reference state model. To obtain these, we perform a cubic spline interpolation to the output of the MESA models and used the Hann function as a window to smooth all abrupt variations in the profiles. Furthermore, we clip the profiles between 1\% and 90\% of the total stellar radius (R$_{\mathrm{star}}$) to avoid numerical instabilities due to the large density stratification near the stellar surface. The middle panel displays the square of \bvf{} or the buoyancy frequency, an output from MESA used to calculate $d\overline{S}/dr$ (see Eq.~(2)) in the radiation zone. The inclusion of the \bvf{} spike at the convective-radiative interface, left over from the receding convective core, has also not been investigated in previous studies.  These are crucial differences that allow these simulations to better inform asteroseismology, which is extremely sensitive to this region \citep{Bowman2020c}.

The bottom panel of Fig.~\ref{fig:stellar_param}, shows the different values of the diffusivities and viscosity used in our MHD simulations. In the convection zone, we have used a much higher thermal diffusivity, at $7 \times 10^{12}$ cm$^2$ s$^{-1}$, shown by the green line, than the  stellar thermal diffusivity generated by MESA, shown by the red line. This is also true for the viscosity, at also $7 \times 10^{12}$ cm$^2$ s$^{-1}$, and the magnetic diffusivity, at $2.5 \times 10^{12}$ cm$^2$ s$^{-1}$, which was done to ensure numerical stability. At the convective-radiative boundary, we decrease the diffusivities since they are meant to mimic ``turbulent'' diffusivities and would be physically smaller in the radiative region. We reduce $\overline{\kappa}$ to match the realistic thermal diffusivity profiles towards the top of the domain. The viscosity and magnetic diffusivity profiles are then decreased to match the thermal diffusivity profiles up to approximately 0.592 R$_{\mathrm{star}}$, where both the viscosity and magnetic diffusivity profiles are set to a constant value of $1 \times 10^{11}$ cm$^2$ s$^{-1}$ towards the top of the domain. Table~\ref{tab:param} provides a summary of the parameter space of our model.

\begin{table}[ht!]
\hskip-0.5cm
\resizebox{\columnwidth}{!}{
\begin{tabular}{ccc}
\hline\hline
 Parameters & \multicolumn{2}{c}{Values (in cgs units)} \\
\hline
 R$_{\mathrm{domain}}$/cm & \multicolumn{2}{c}{3.04 $\times 10^{11}$} \\
 R$_{\mathrm{CZ}}$/cm & \multicolumn{2}{c}{4.04 $\times 10^{10}$} \\
 $\tau_{\mathrm{rot}}$/days & \multicolumn{2}{c}{4.04} \\
  $\tau_{\mathrm{sim}}$/days & \multicolumn{2}{c}{275} \\
 c$_s$/cm s$^{-1}$ & \multicolumn{2}{c}{$2.0\times 10^{6}$ (minimum)} \\
 \hline
       & CZ & RZ \\
\hline
 v$_{\mathrm{rms}}$/cm s$^{-1}$ & 82300 cm s$^{-1}$ & 1130 cm s$^{-1}$ \\
 Re & 823    \\
 Re$_{\mathrm{m}}$ & 1320    \\
 Ro & 0.718  \\
 $\tau_{\kappa}$/days & 3090 & 14400/1.18 $\times 10^9$ \\
 $\tau_{\eta}$/days & 7410 & 7.82$\times 10^7$/1.18 $\times 10^9$ \\
 $\tau_{\mathrm{CZ}}$/days & 6.62 \\
 \hline\hline
\end{tabular}
}
\caption{Stellar model parameters. The radius of the simulation domain and the convection zone are represented by R$_{\mathrm{domain}}$ and R$_{\mathrm{CZ}}$, respectively. $\tau_{\mathrm{rot}}$ and $\tau_{\mathrm{sim}}$ are the rotational period and total simulation time, whilst $c_{s}$ is minimum sound speed in the whole domain. Under the convection zone (CZ) and radiation zone (RZ) headers, v$_{\mathrm{rms}}$ is the root-mean-square velocity. The Reynold's and magnetic Reynold's numbers are represented by Re and Re$_{\mathrm{m}}$, while the Rossby number by Ro. The thermal diffusion timescale is $\tau_{\kappa} = L^2/\kappa$ and the magnetic diffusion timescale is $\tau_{\eta} = L^2/\eta$, where L is either the size of the convection zone or the radiation zone. $\tau_{\mathrm{CZ}}$ is the convective timescale and is calculated as R$_{\mathrm{CZ}}$/v$_{\mathrm{rms}}$. \label{tab:param}}
\end{table} 


\section{Magnetic Field Evolution}\label{sec:mag_field_evo}

\begin{figure}[ht!]
	\begin{minipage}{\columnwidth}
		\includegraphics[width=\columnwidth]{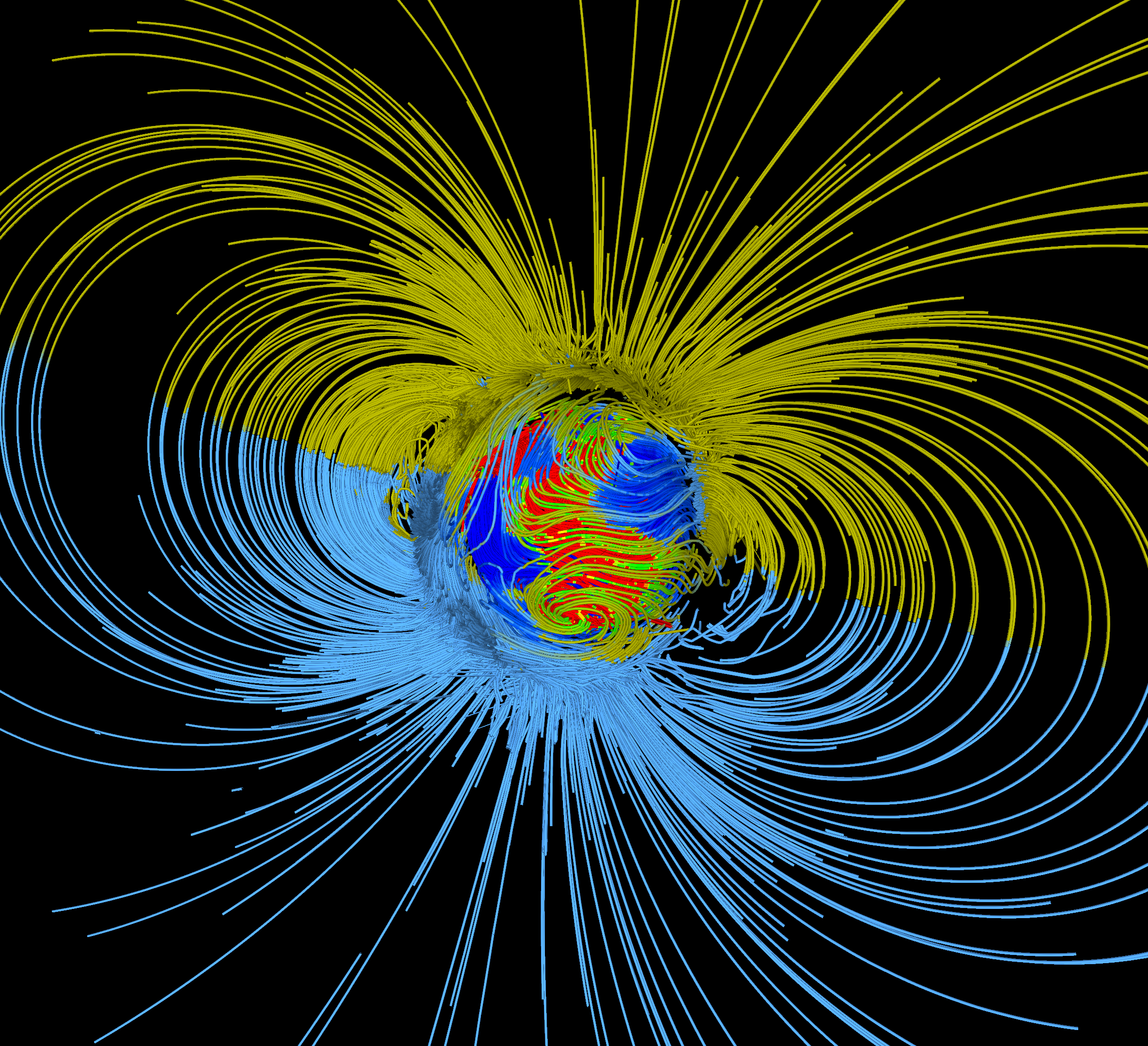}
	\end{minipage}
	\caption{Magnetic field geometry from our simulations. The gold-blue magnetic field lines show a predominantly dipolar field geometry within the radiative envelope, whereas the dark blue-red-green magnetic field lines show a more complex magnetic field structure inside the convective core.}
	\label{fig:vtests}
\end{figure}

We begin by solving only the hydrodynamical equations to allow the onset of stellar convection until steady-state velocities are reached. Once the simulation reaches this stage, we introduce a weak dipolar magnetic field of strength $\sim$ 1~G and solve the full MHD equations. The interaction between the convective motions and the magnetic field leads to large increases in the magnetic energy in the convective core as the dynamo is established. In the radiative zone the large scale dipole field decays on a magnetic diffusion timescale, which is much larger than our total simulation time (see $\tau_{\mathrm{sim}}$ and $\tau_{\eta}$ in Table~\ref{tab:param}).

Figure~\ref{fig:vtests} shows the magnetic streamlines when the magnetic field is in steady-state, where we observe a complex magnetic field structure in the core, surrounded by a dipolar field in the radiation zone. The magnetic field inside the core is sustained by a magnetic dynamo, generated by the convective motions. Starting from the convective-radiative boundary, where the \bvf{} frequency is zero (see Fig.~\ref{fig:stellar_param}), the magnetic field strength decreases with radius until it matches the imposed large-scale dipolar magnetic field. However, close to the boundary, Fig.~\ref{fig:vtests} shows that the magnetic field is dominated by streamlines in the azimuthal direction, representative of a strong toroidal field in that region.

\section{Near-Core Magnetic Field Geometry}\label{sec:interaction}
\begin{figure*}[ht!]
        \centering
        \includegraphics[trim={0.cm 0.cm 0.cm 0.0cm},clip,width=\textwidth]{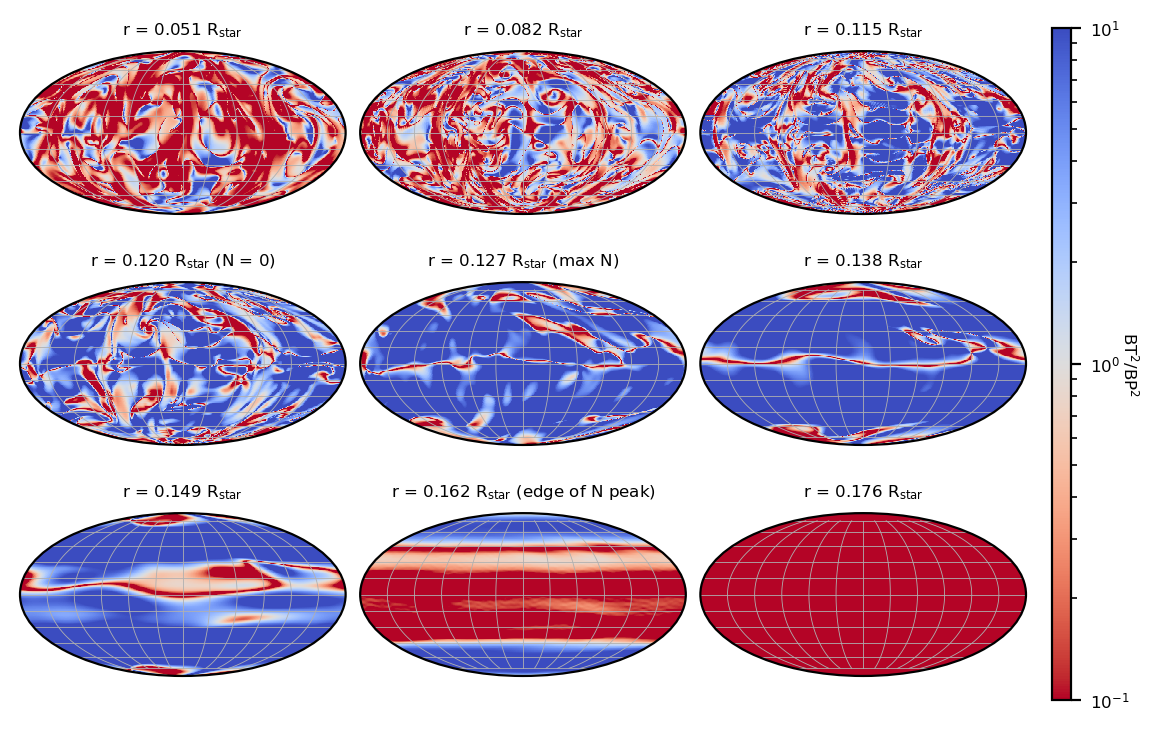}
        \caption{Mollweide plots showing the ratio of the toroidal magnetic field energy to poloidal field energy, $\mathrm{BT}^2/\mathrm{BP}^2$ as functions of latitudes and longitudes over a sphere. The plots in the top row show the ratios inside the convection zone. At 0.12 R$_{\mathrm{star}}$, the \bvf{} frequency is approximately 0 Hz. The remaining plots show the ratios inside the \bvf{} frequency peak from its maximum value at 0.127 R$_{\mathrm{star}}$ to just outside the peak at 0.176 R$_{\mathrm{star}}$ (see Fig.~\ref{fig:stellar_param}). \label{fig:mollweide_snapshot}}
        \centering
\end{figure*}


The main aim of this work is to determine the magnetic field geometry at the core-envelope boundary of a mid-main sequence massive star, so we start with a time snapshot of the magnetic field configuration at different shells (radii) after the system has evolved (see Fig.~\ref{fig:mollweide_snapshot}). This time was chosen to be within the steady-state evolution of both the averaged kinetic and magnetic energies inside the convection zone, which were reached within 30 days for the kinetic energy and 50 days for the magnetic energy. At this point, the average magnetic field is around 100 kG in the convection zone and lower in the shear layer, dropping to the imposed 10 G at the top of the shear layer. Note that this is within the upper limit constraint of 500 kG placed on HD~43317. We show the ratio of the toroidal to poloidal components of the magnetic field, mainly to demonstrate the changes to the field geometry, using Mollweide projection plots in this figure. The first row shows that there is more energy in the poloidal components of the magnetic field compared to the energy in the toroidal component deeper inside the convection core. As the radius approaches 0.12 R$_{\mathrm{star}}$, which is the location where $N = 0$, we observe more energy in the toroidal magnetic field. Moving past this boundary into the \bvf{} frequency peak, we observe a larger surface area of the shell, having a ratio of toroidal to poloidal energies that exceeds unity. At r = 0.127 R$_{\mathrm{star}}$, the ratio of toroidal to poloidal field peaks with $B_T/B_P > 10$ almost over the whole shell. This is true up to r = 0.149 R$_{\mathrm{star}}$ and at radii larger than this, the poloidal field starts to dominate again, as the ratio of energies starts to fall below unity over the whole shell. 

\begin{figure*}[ht!]
        \centering
        \includegraphics[trim={0.0cm 0.cm 0.cm 0.cm},clip,width=\textwidth]{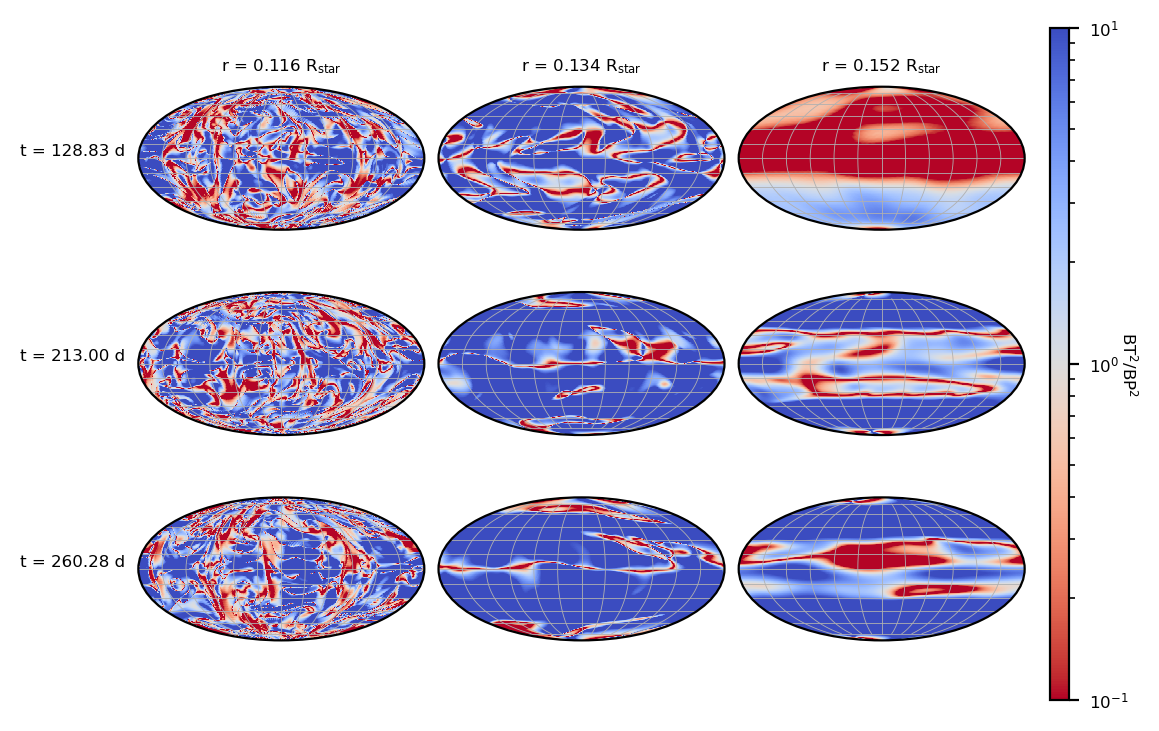}
        \caption{The ratios of the toroidal magnetic field energy to poloidal field energy, $\mathrm{BT}^2/\mathrm{BP}^2$ as functions of latitudes and longitudes at different times (rows) and different radii (columns). \label{fig:mollweide_time}}
        \centering
\end{figure*}

\begin{figure}[ht!]
        \centering
        \includegraphics[trim={0.0cm 0.0cm 0 0.0cm},clip,width=\columnwidth]{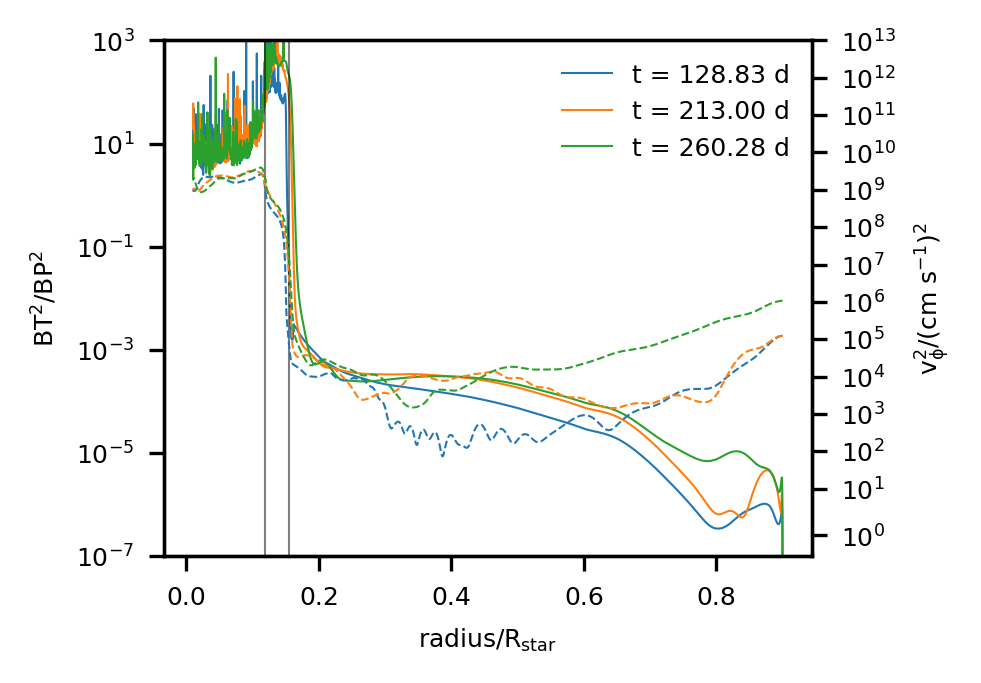}
        \caption{The shell-averaged ratios of the toroidal magnetic field energy to poloidal field energy, $\mathrm{BT}^2/\mathrm{BP}^2$ (solid lines) and azimuthal velocities (dashed lines) as a function of radius. The different colours represent different points in time during the simulation. The solid, grey vertical lines represent the extent of the \bvf{} frequency peak.\label{fig:line_ratio}}
        \centering
\end{figure}

The randomly-spread poloidal and toroidal energies inside the convection core generally follow the pattern of convective fluid motions. However, around the convective-radiative boundary, a strong shear layer forms within the \bvf{} frequency peak, perpendicular to the stellar rotational axis (see Figure 5). The rotation profile is columnar, exhibiting a strong radial differential rotation at the interface, with prograde flow at lower latitudes and retrograde motion at higher latitudes. The radial shear drags magnetic field lines in the azimuthal direction, causing the toroidal magnetic field to become stronger within the shear layer \citep{PittsTayler1985,Zahn2007}. The latitudinal differential rotation causes the toroidal field to change sign at mid-latitudes. Figure~\ref{fig:mollweide_time} shows the Mollweide projections at different times as different rows, and for different radii as different columns. In the first column, where the shells are located just below the convective-radiative boundary, there are no large differences in the magnetic field energy distribution between the toroidal and poloidal components. As we move into the shear layer within the \bvf{} peak, the toroidal field energy begins to dominate over time at all latitudes. Further inside the \bvf{} frequency peak, regions that are initially dominated by the poloidal field can be seen to become to more dominated by the toroidal field, away from the equatorial plane. 


In Fig.~\ref{fig:line_ratio}, we show the shell-averaged radial profiles of the ratio of toroidal to poloidal magnetic energies (solid lines), alongside the azimuthal velocity squared, relative to the rotating frame (dashed lines) at different times. Within the convection zone, the ratios range from 10-100, but vary dramatically as seen in Fig.~\ref{fig:mollweide_time}.  Moving into the convective-radiative interface, demarcated by the \bvf{} frequency peak, the toroidal field dominates throughout. The toroidal field is generally 100 times stronger than the poloidal field and maps regions of strong azimuthal shear, as expected. Outside this region, within the bulk of the radiation zone, the (seed) poloidal field dominates.  

Within the shear layer, we also find that, at steady-state evolution, the ratio of the magnetic energy to the kinetic energy ranges between 0.15 and 0.4. These ratios are fairly low, and hence the magnetic field strength is not large enough to reduce the hydrodynamically-driven differential rotation in the region. However, at larger rotation rates, where a larger field might be induced, the near-core differential rotation could be reduced by magnetic tension.

\section{Conclusions} \label{sec:conclusion}

We present 3D spherical MHD simulations of the internal fluid and magnetic field evolution of a mid-main sequence 7~\msol{} star at 11.5\% the critical breakup velocity. The simulations were run with RAYLEIGH using reference state data from the 1D stellar evolution code MESA. The simulation domain covers between 1\% to 90\% of the total stellar radius, and the initial magnetic seed was set as a weak dipole field. 

We find that the geometry of the magnetic field evolves and changes from its original dipolar configuration to a more complex configuration inside the convection zone, due to the action of the stellar dynamo, whilst remaining unaffected in the radiation zone. At the convective-radiative interface, the ratio of toroidal to poloidal field increases with time, which is driven by the shear between the regions. We also observe a latitudinal dependence consistent with the same shear flow that coincides exactly with the peak in the \bvf{} frequency profile. These results apply to specific stellar and MHD parameters (e.g. Re$_{\mathrm{m}}$ and initial rotation rates).  A faster initial rotation or a larger Reynolds number could both lead to a stronger magnetic field, which could reduce differential rotation across the BVF peak. Therefore, a broader parameter study is required to ascertain the variety of behaviour possible. 

The \bvf{} frequency peak just outside the convective core is precisely the region that asteroseismology of pulsation modes in SPB stars is most sensitive to \citep{Bowman2020c}. Not many main-sequence stars born with a convective core have been investigated for the amount of differential radial rotation within the \bvf{} frequency peak (e.g. \citealt{Aerts2003d, VanReeth2018a, Burssens2023a}), but such work generally requires the extent and shape of the shear layer to be assumed. Recently, \citet{Burssens2023a} was the first to test the impact of this assumption on the resultant differential rotation profiles for the mid-main sequence 12M$_{\odot}$ star HD~192575. They demonstrated the significant impact on the inferred core-to-envelope rotation ratio when assuming the shear layer to be either the width of the \bvf{} frequency or the (much) smaller CBM region. The former was shown to provide the statistically better fit, which is supported by 2D structure models by \citet{Mombarg2023b} and our 3D MHD simulations presented in this work.

To date, only a single asteroseismic study has probed the properties of an interior magnetic field deep inside an SPB star. Assuming a purely dipolar field geometry and a rigid rotation profile for the mid-main sequence SPB star HD~43317 \citep{Buysschaert2018c}, an upper limit of the near-core magnetic field was constrained to be $5 \times 10^5$~G \citep{Lecoanet2022a}. Furthermore, there are currently no forward modelling results that directly include magnetic fields. Parameter studies of massive stars that include the impact of a magnetic field include \cite{Prat2019,Prat2020} and \cite{van_beeck_2020}, which show significant differences in pulsation mode frequencies when using a perturbative approach with a mixed poloidal and toroidal magnetic field topology. In our work, we provide numerical evidence that the magnetic field topology for such stars in the near-core region is predominantly toroidal. Moreover, we demonstrate that the location of the shear layer and the predominantly toroidal magnetic field geometry is within the \bvf{} frequency peak for a representative mid-main sequence massive star. These results are important for future forward-asteroseismic modelling studies that aim to constrain interior rotation rates, mixing profiles, and interior magnetic fields for stars born with convective cores.

\vspace{0.2cm}
{\it Acknowledgements:} 
We would like to thank the referee for providing useful feedback that helped us improve our work. We acknowledge support from STFC grant ST/W001020/1. Computing was carried out on the DiRAC Data Intensive service at Leicester (DIaL), operated by the University of Leicester IT Services, which forms part of the STFC DiRAC HPC Facility (\url{www.dirac.ac.uk}), funded by BEIS capital funding via STFC capital grants ST/K000373/1 and ST/R002363/1 and STFC DiRAC Operations grant ST/R001014/1. PVFE was supported by the U.S. Department of Energy through the Los Alamos National Laboratory (LANL). LANL is operated by Triad National Security, LLC, for the National Nuclear Security Administration of the U.S. Department of Energy (Contract No. 89233218CNA000001). DMB gratefully acknowledges funding from UK Research and Innovation (UKRI) in the form of a Frontier Research Grant under the UK government's ERC Horizon Europe funding guarantee (SYMPHONY; grant number: EP/Y031059/1), and a Royal Society University Research Fellowship (URF; grant number: URF{\textbackslash}R1{\textbackslash}231631). This work has been assigned a document release number LA-UR-24-25321.

\bibliographystyle{aasjournal}
\bibliography{paper4}{}

\newcommand{\noop}[1]{}
\begin{thebibliography}{}
\expandafter\ifx\csname natexlab\endcsname\relax\def\natexlab#1{#1}\fi
\providecommand{\url}[1]{\href{#1}{#1}}
\providecommand{\dodoi}[1]{doi:~\href{http://doi.org/#1}{\nolinkurl{#1}}}
\providecommand{\doeprint}[1]{\href{http://ascl.net/#1}{\nolinkurl{http://ascl.net/#1}}}
\providecommand{\doarXiv}[1]{\href{https://arxiv.org/abs/#1}{\nolinkurl{https://arxiv.org/abs/#1}}}

\bibitem[{{Aerts}(2021)}]{Aerts2021}
{Aerts}, C. 2021, Reviews of Modern Physics, 93, 015001, \dodoi{10.1103/RevModPhys.93.015001}

\bibitem[{Aerts {et~al.}(2010)Aerts, Christensen-Dalsgaard, \& Kurtz}]{aerts2010asteroseismology}
Aerts, C., Christensen-Dalsgaard, J., \& Kurtz, D. 2010, Asteroseismology, Astronomy and Astrophysics Library (Springer Netherlands).
\newblock \url{https://books.google.co.uk/books?id=N8pswDrdSyUC}

\bibitem[{{Aerts} {et~al.}(2003){Aerts}, {Thoul}, {Daszy{\'n}ska}, {Scuflaire}, {Waelkens}, {Dupret}, {Niemczura}, \& {Noels}}]{Aerts2003d}
{Aerts}, C., {Thoul}, A., {Daszy{\'n}ska}, J., {et~al.} 2003, Science, 300, 1926, \dodoi{10.1126/science.1084993}

\bibitem[{{Augustson} {et~al.}(2016){Augustson}, {Brun}, \& {Toomre}}]{Auguston2016}
{Augustson}, K.~C., {Brun}, A.~S., \& {Toomre}, J. 2016, \apj, 829, 92, \dodoi{10.3847/0004-637X/829/2/92}

\bibitem[{{Augustson} {et~al.}(2019){Augustson}, {Brun}, \& {Toomre}}]{Auguston2019}
---. 2019, \apj, 876, 83, \dodoi{10.3847/1538-4357/ab14ea}

\bibitem[{{Bowman}(2020)}]{Bowman2020c}
{Bowman}, D.~M. 2020, Frontiers in Astronomy and Space Sciences, 7, 70, \dodoi{10.3389/fspas.2020.578584}

\bibitem[{{Bowman}(2023)}]{Bowman2023b}
---. 2023, \apss, 368, 107, \dodoi{10.1007/s10509-023-04262-7}

\bibitem[{{Briquet} {et~al.}(2013){Briquet}, {Neiner}, {Leroy}, {P{\'a}pics}, \& {MiMeS Collaboration}}]{Briquet2013}
{Briquet}, M., {Neiner}, C., {Leroy}, B., {P{\'a}pics}, P.~I., \& {MiMeS Collaboration}. 2013, \aap, 557, L16, \dodoi{10.1051/0004-6361/201321779}

\bibitem[{{Briquet} {et~al.}(2012){Briquet}, {Neiner}, {Aerts}, {Morel}, {Mathis}, {Reese}, {Lehmann}, {Costero}, {Echevarria}, {Handler}, {Kambe}, {Hirata}, {Masuda}, {Wright}, {Yang}, {Pintado}, {Mkrtichian}, {Lee}, {Han}, {Bruch}, {De Cat}, {Uytterhoeven}, {Lefever}, {Vanautgaerden}, {de Batz}, {Fr{\'e}mat}, {Henrichs}, {Geers}, {Martayan}, {Hubert}, {Thizy}, \& {Tijani}}]{Briquet2012}
{Briquet}, M., {Neiner}, C., {Aerts}, C., {et~al.} 2012, \mnras, 427, 483, \dodoi{10.1111/j.1365-2966.2012.21933.x}

\bibitem[{{Browning}(2008)}]{Browning2008}
{Browning}, M.~K. 2008, \apj, 676, 1262, \dodoi{10.1086/527432}

\bibitem[{{Brun} {et~al.}(2005){Brun}, {Browning}, \& {Toomre}}]{Brun2005}
{Brun}, A.~S., {Browning}, M.~K., \& {Toomre}, J. 2005, \apj, 629, 461, \dodoi{10.1086/430430}

\bibitem[{Burns {et~al.}(2020)Burns, Vasil, Oishi, Lecoanet, \& Brown}]{Burns2020}
Burns, K.~J., Vasil, G.~M., Oishi, J.~S., Lecoanet, D., \& Brown, B.~P. 2020, Phys. Rev. Research, 2, 023068, \dodoi{10.1103/PhysRevResearch.2.023068}

\bibitem[{{Burssens} {et~al.}(2023){Burssens}, {Bowman}, {Michielsen}, {Sim{\'o}n-D{\'\i}az}, {Aerts}, {Vanlaer}, {Banyard}, {Nardetto}, {Townsend}, {Handler}, {Mombarg}, {Vanderspek}, \& {Ricker}}]{Burssens2023a}
{Burssens}, S., {Bowman}, D.~M., {Michielsen}, M., {et~al.} 2023, Nature Astronomy, 7, 913, \dodoi{10.1038/s41550-023-01978-y}

\bibitem[{{Buysschaert} {et~al.}(2018){Buysschaert}, {Aerts}, {Bowman}, {Johnston}, {Van Reeth}, {Pedersen}, {Mathis}, \& {Neiner}}]{Buysschaert2018c}
{Buysschaert}, B., {Aerts}, C., {Bowman}, D.~M., {et~al.} 2018, \aap, 616, A148, \dodoi{10.1051/0004-6361/201832642}

\bibitem[{{Buysschaert} {et~al.}(2017){Buysschaert}, {Neiner}, {Briquet}, \& {Aerts}}]{Buysschaert2017b}
{Buysschaert}, B., {Neiner}, C., {Briquet}, M., \& {Aerts}, C. 2017, \aap, 605, A104, \dodoi{10.1051/0004-6361/201731012}

\bibitem[{{Charbonneau} \& {Sokoloff}(2023)}]{Charbonneau2023}
{Charbonneau}, P., \& {Sokoloff}, D. 2023, \ssr, 219, 35, \dodoi{10.1007/s11214-023-00980-0}

\bibitem[{Featherstone \& Hindman(2016)}]{featherstone2022paper}
Featherstone, N., \& Hindman, B. 2016, The Astrophysical Journal, 818, 32, \dodoi{http://doi.org/10.3847/0004-637X/818/1/32}

\bibitem[{{Featherstone} {et~al.}(2009){Featherstone}, {Browning}, {Brun}, \& {Toomre}}]{Featherstone2009}
{Featherstone}, N.~A., {Browning}, M.~K., {Brun}, A.~S., \& {Toomre}, J. 2009, \apj, 705, 1000, \dodoi{10.1088/0004-637X/705/1/1000}

\bibitem[{{Featherstone} {et~al.}(2011){Featherstone}, {Browning}, {Brun}, \& {Toomre}}]{Featherstone2011}
{Featherstone}, N.~A., {Browning}, M.~K., {Brun}, A.~S., \& {Toomre}, J. 2011, in Journal of Physics Conference Series, Vol. 271, GONG-SoHO 24: A New Era of Seismology of the Sun and Solar-Like Stars (IOP), 012068, \dodoi{10.1088/1742-6596/271/1/012068}

\bibitem[{{Featherstone} {et~al.}(2022){Featherstone}, {Edelmann}, {Gassmoeller}, {Matilsky}, {Orvedahl}, \& {Wilson}}]{featherstone_et_al_2022}
{Featherstone}, N.~A., {Edelmann}, P.~V.~F., {Gassmoeller}, R., {et~al.} 2022, Rayleigh 1.1.0, \dodoi{http://doi.org/10.5281/zenodo.6522806}

\bibitem[{{Lecoanet} {et~al.}(2022){Lecoanet}, {Bowman}, \& {Van Reeth}}]{Lecoanet2022a}
{Lecoanet}, D., {Bowman}, D.~M., \& {Van Reeth}, T. 2022, \mnras, 512, L16, \dodoi{10.1093/mnrasl/slac013}

\bibitem[{Matsui {et~al.}(2016)Matsui, Heien, Aubert, Aurnou, Avery, Brown, Buffett, Busse, Christensen, Davies, Featherstone, Gastine, Glatzmaier, Gubbins, Guermond, Hayashi, Hollerbach, Hwang, Jackson, Jones, Jiang, Kellogg, Kuang, Landeau, Marti, Olson, Ribeiro, Sasaki, Schaeffer, Simitev, Sheyko, Silva, Stanley, Takahashi, Takehiro, Wicht, \& Willis}]{Matsui_etal_2016}
Matsui, H., Heien, E., Aubert, J., {et~al.} 2016, Geochemistry, Geophysics, Geosystems, 17, 1586, \dodoi{http://doi.org/10.1002/2015GC006159}

\bibitem[{{Michielsen} {et~al.}(2021){Michielsen}, {Aerts}, \& {Bowman}}]{Michielsen2021a}
{Michielsen}, M., {Aerts}, C., \& {Bowman}, D.~M. 2021, \aap, 650, A175, \dodoi{10.1051/0004-6361/202039926}

\bibitem[{{Mombarg} {et~al.}(2023){Mombarg}, {Rieutord}, \& {Espinosa Lara}}]{Mombarg2023b}
{Mombarg}, J.~S.~G., {Rieutord}, M., \& {Espinosa Lara}, F. 2023, \aap, 677, L5, \dodoi{10.1051/0004-6361/202347454}

\bibitem[{{Moravveji} {et~al.}(2015){Moravveji}, {Aerts}, {P{\'a}pics}, {Triana}, \& {Vandoren}}]{Moravveji2015b}
{Moravveji}, E., {Aerts}, C., {P{\'a}pics}, P.~I., {Triana}, S.~A., \& {Vandoren}, B. 2015, \aap, 580, A27, \dodoi{10.1051/0004-6361/201425290}

\bibitem[{{Moravveji} {et~al.}(2016){Moravveji}, {Townsend}, {Aerts}, \& {Mathis}}]{Moravveji2016b}
{Moravveji}, E., {Townsend}, R.~H.~D., {Aerts}, C., \& {Mathis}, S. 2016, \apj, 823, 130, \dodoi{10.3847/0004-637X/823/2/130}

\bibitem[{{P{\'a}pics} {et~al.}(2012){P{\'a}pics}, {Briquet}, {Baglin}, {Poretti}, {Aerts}, {Degroote}, {Tkachenko}, {Morel}, {Zima}, {Niemczura}, {Rainer}, {Hareter}, {Baudin}, {Catala}, {Michel}, {Samadi}, \& {Auvergne}}]{Papics2012a}
{P{\'a}pics}, P.~I., {Briquet}, M., {Baglin}, A., {et~al.} 2012, \aap, 542, A55, \dodoi{10.1051/0004-6361/201218809}

\bibitem[{{P{\'a}pics} {et~al.}(2017){P{\'a}pics}, {Tkachenko}, {Van Reeth}, {Aerts}, {Moravveji}, {Van de Sande}, {De Smedt}, {Bloemen}, {Southworth}, {Debosscher}, {Niemczura}, \& {Gameiro}}]{Papics2017a}
{P{\'a}pics}, P.~I., {Tkachenko}, A., {Van Reeth}, T., {et~al.} 2017, \aap, 598, A74, \dodoi{10.1051/0004-6361/201629814}

\bibitem[{{Paxton} {et~al.}(2011){Paxton}, {Bildsten}, {Dotter}, {Herwig}, {Lesaffre}, \& {Timmes}}]{mesa_1}
{Paxton}, B., {Bildsten}, L., {Dotter}, A., {et~al.} 2011, \apjs, 192, 3, \dodoi{10.1088/0067-0049/192/1/3}

\bibitem[{{Paxton} {et~al.}(2013){Paxton}, {Cantiello}, {Arras}, {Bildsten}, {Brown}, {Dotter}, {Mankovich}, {Montgomery}, {Stello}, {Timmes}, \& {Townsend}}]{mesa_2}
{Paxton}, B., {Cantiello}, M., {Arras}, P., {et~al.} 2013, \apjs, 208, 4, \dodoi{10.1088/0067-0049/208/1/4}

\bibitem[{Paxton {et~al.}(2015)Paxton, Marchant, Schwab, Bauer, Bildsten, Cantiello, Dessart, Farmer, Hu, Langer, Townsend, Townsley, \& Timmes}]{mesa_3}
Paxton, B., Marchant, P., Schwab, J., {et~al.} 2015, \apjs, 220, 15, \dodoi{10.1088/0067-0049/220/1/15}

\bibitem[{{Paxton} {et~al.}(2018){Paxton}, {Schwab}, {Bauer}, {Bildsten}, {Blinnikov}, {Duffell}, {Farmer}, {Goldberg}, {Marchant}, {Sorokina}, {Thoul}, {Townsend}, \& {Timmes}}]{mesa_4}
{Paxton}, B., {Schwab}, J., {Bauer}, E.~B., {et~al.} 2018, \apjs, 234, 34, \dodoi{10.3847/1538-4365/aaa5a8}

\bibitem[{{Paxton} {et~al.}(2019){Paxton}, {Smolec}, {Schwab}, {Gautschy}, {Bildsten}, {Cantiello}, {Dotter}, {Farmer}, {Goldberg}, {Jermyn}, {Kanbur}, {Marchant}, {Thoul}, {Townsend}, {Wolf}, {Zhang}, \& {Timmes}}]{mesa_5}
{Paxton}, B., {Smolec}, R., {Schwab}, J., {et~al.} 2019, \apjs, 243, 10, \dodoi{10.3847/1538-4365/ab2241}

\bibitem[{{Pedersen} {et~al.}(2021){Pedersen}, {Aerts}, {P{\'a}pics}, {Michielsen}, {Gebruers}, {Rogers}, {Molenberghs}, {Burssens}, {Garcia}, \& {Bowman}}]{Pedersen2021a}
{Pedersen}, M.~G., {Aerts}, C., {P{\'a}pics}, P.~I., {et~al.} 2021, Nature Astronomy, 5, 715, \dodoi{10.1038/s41550-021-01351-x}

\bibitem[{Pitts \& Tayler(1985)}]{PittsTayler1985}
Pitts, E., \& Tayler, R.~J. 1985, Monthly Notices of the Royal Astronomical Society, 216, 139, \dodoi{10.1093/mnras/216.2.139}

\bibitem[{{Prat} {et~al.}(2019){Prat}, {Mathis}, {Buysschaert}, {Van Beeck}, {Bowman}, {Aerts}, \& {Neiner}}]{Prat2019}
{Prat}, V., {Mathis}, S., {Buysschaert}, B., {et~al.} 2019, A\&A, 627, A64, \dodoi{10.1051/0004-6361/201935462}

\bibitem[{{Prat} {et~al.}(2020){Prat}, {Mathis}, {Neiner}, {Van Beeck}, {Bowman}, \& {Aerts}}]{Prat2020}
{Prat}, V., {Mathis}, S., {Neiner}, C., {et~al.} 2020, A\&A, 636, A100, \dodoi{10.1051/0004-6361/201937398}

\bibitem[{{Spruit}(1999)}]{Spruit1999}
{Spruit}, H.~C. 1999, \aap, 349, 189, \dodoi{10.48550/arXiv.astro-ph/9907138}

\bibitem[{{Szewczuk} {et~al.}(2021){Szewczuk}, {Walczak}, \& {Daszy{\'n}ska-Daszkiewicz}}]{Szewczuk2021a}
{Szewczuk}, W., {Walczak}, P., \& {Daszy{\'n}ska-Daszkiewicz}, J. 2021, \mnras, 503, 5894, \dodoi{10.1093/mnras/stab683}

\bibitem[{{Van Beeck} {et~al.}(2020){Van Beeck}, {Prat}, {Van Reeth}, {Mathis}, {Bowman}, {Neiner}, \& {Aerts}}]{van_beeck_2020}
{Van Beeck}, J., {Prat}, V., {Van Reeth}, T., {et~al.} 2020, A\&A, 638, A149, \dodoi{10.1051/0004-6361/201937363}

\bibitem[{{Van Reeth} {et~al.}(2018){Van Reeth}, {Mombarg}, {Mathis}, {Tkachenko}, {Fuller}, {Bowman}, {Buysschaert}, {Johnston}, {Garc{\'{\i}}a Hern{\'a}ndez}, {Goldstein}, {Townsend}, \& {Aerts}}]{VanReeth2018a}
{Van Reeth}, T., {Mombarg}, J.~S.~G., {Mathis}, S., {et~al.} 2018, \aap, 618, A24, \dodoi{10.1051/0004-6361/201832718}

\bibitem[{{Zahn} {et~al.}(2007){Zahn}, {Brun}, \& {Mathis}}]{Zahn2007}
{Zahn}, J.~P., {Brun}, A.~S., \& {Mathis}, S. 2007, \aap, 474, 145, \dodoi{10.1051/0004-6361:20077653}

\end{thebibliography}



\end{document}